\documentclass[aps,prb,superscriptaddress,twocolumn]{revtex4}
\usepackage[usenames]{color}
\usepackage{graphicx}
\begin{document}
\title{Local distortions in slow moving vortex lattices}
\author{Michael Dreyer}
\email{dreyer@lps.umd.edu}
\author{Jonghee Lee}
\author{Hui Wang}
\affiliation{University of Maryland, College Park, MD 20740.}
\affiliation{Laboratory for Physical Sciences, 8050 Greenmead Drive, College Park, MD 20740.}
\author{Barry Barker}
\affiliation{Laboratory for Physical Sciences, 8050 Greenmead Drive, College Park, MD 20740.}
\date{\today}

\begin{abstract}
We observed slow moving vortex lattices in NbSe$_2$ using a home built STM at a temperature of 4.2~K and magnetic fields in the range of $250-500$~mT. The vortex lattices move coherently at speeds of $1-10$~pm/s due to a slow decay of our magnetic field of 0.5~mT/day. We observe collective speed variations which indicate pinning/depinning events. Furthermore, we found -- to our knowledge for the first time -- small and  local deviations from the coherent behavior. Most noticeable were local lattice distortions with displacements of the vortices of $\sim 2$~nm from their ideal lattice position. We compared the observations to a 2-D simulation of a vortex lattice moving under similar conditions and found similar distortion effects near pinning sites. Thus, studying the distortion patterns could lead to a better understanding of the vortex--defect interaction.
\end{abstract}

\maketitle

\section{Introduction}

NbSe$_2$ is a widely studied material. It is regularly used as a test sample for low temperature scanning tunneling microscopes operating in magnetic fields since it is easy to prepare and shows a charge density wave. In addition it is a type II superconductor allowing the calibration of the magnetic field by observing the vortex pattern. At low magnetic fields pinning centers, such as structural defects in the NbSe$_2$ crystal get populated. Towards higher fields the Abrikosov lattice\cite{abrikosov_lat} is formed. For our conditions of field and temperature the lattice should be in the Bragg glass phase\cite{giamarchi_bragg,giamarchi_bragg2} with local hexagonal but no long range order. It has also been previously reported that the vortices move under the influence of a changing magnetic field. The observations were performed using Bitter technique\cite{pardo_bishop_evlm} at low fields, using muon scattering\cite{sonier_kiefl_emu_sr}, as well as using STM at higher fields\cite{troyanovski_kes_evlm}. However, the measured velocities were typically several orders of magnitude higher than in our study ($\mu$m/s vs.\ pm/s), making local details of the vortex motion harder to observe. In our case, the vortices move about $0.25-2.5$~nm between two passes of the STM tip resulting in a highly resolved time series. This enables us to determine details of the vortex motion unseen in previous experiments.

\begin{figure}%
\includegraphics[width=0.45\columnwidth]{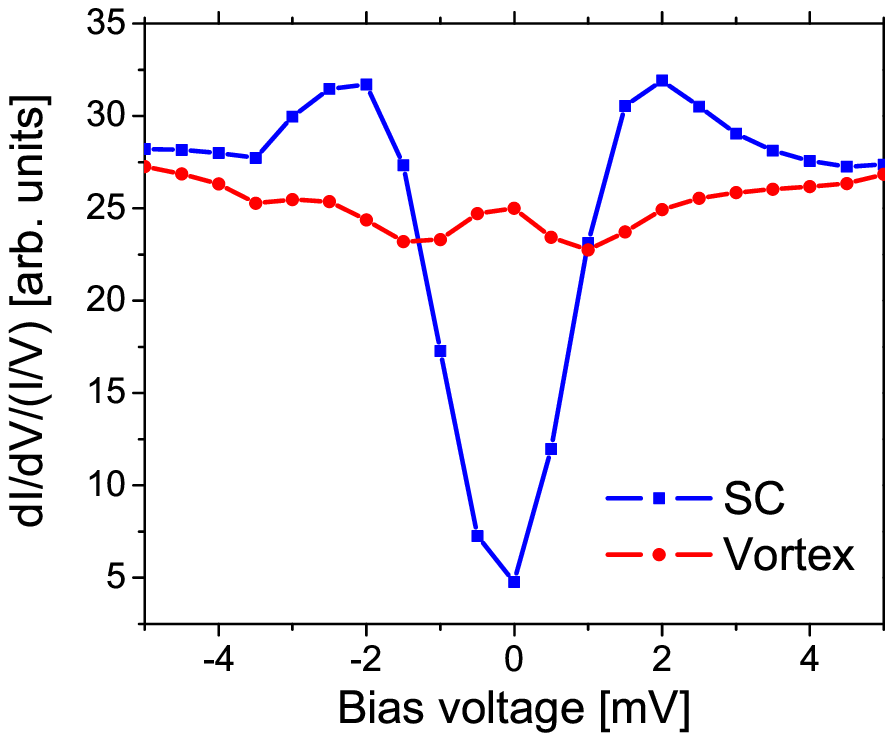}\hfil%
\includegraphics[width=0.35\columnwidth]{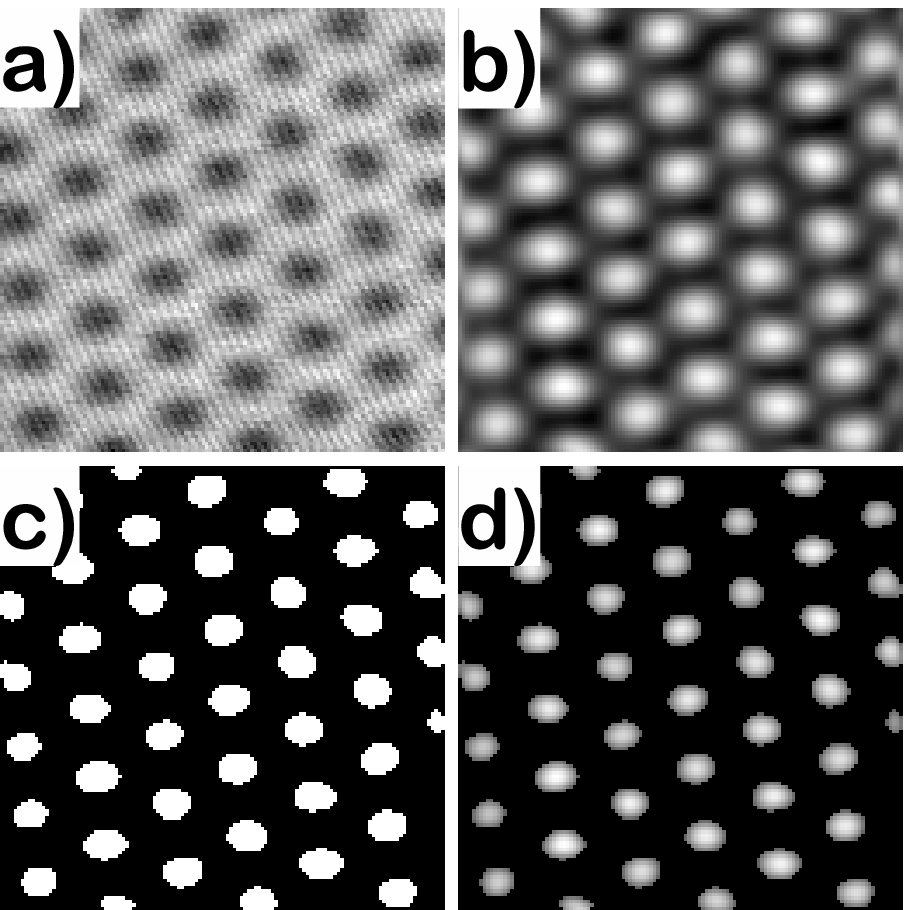}%
\caption{\label{fig:data_proc}(Color online) Normalized dI/dV curves of the superconducting phase and the vortex core (left) and dI/dV image of the vortex phase (400 nm $\times$ 400 nm) in various stages of being processed (right): a) original, b) inverted and filtered, c) threshold to find vortex, d) product of b) and c) to find vortex center as a weighted sum.}%
\end{figure}

As described elsewhere\cite{jl_vm}, consecutive images of the vortex lattice were taken by measuring variations in the local density of states near the superconducting gap (typically: $V_\mathrm{bias}=3$ mV, $I_\mathrm{t}=0.1$ nA). The images were automatically processed to extract the vortex positions with sub-pixel resolution (See Fig.\ \ref{fig:data_proc}). The position of each vortex was followed through the image series to extract the velocity and the vortex lattice was reconstructed to find the local lattice constant. We found collective variations in the vortex velocity which primarily demonstrate the cohesive behavior of the vortex lattice and whose source lies thus lies outside of the observed area. In addition we found local lattice distortions on the order of $\sim 2$ nm which are the main topic of this paper. Detailed analysis of the data revealed that the lattice distortion is also accompanied by local variations in the vortex velocity, which is necessary to maintain the vortex lattice. While clearly visible, these higher order effects suffer a poor signal to noise ratio and shall not be discussed in this paper. The distortions are distinctly different from dislocation (eg.\ recent overview of Bitter decoration\cite{fasano_bitter}) in that the lattice ordering and six fold coordination stays intact. A second distinction is that we are observing a moving vortex matter. Hence, the stationary distortion must be rooted in interactions with the host material. In simulations similar distortion patterns were found near the location of weak pinning centers. While for a detailed study of vortex--defect interactions Ginzburg-Landau calculations similar to works by Maurer et.\ al\cite{maurer_ea_glvpi} are necessary, we chose a 2D Monte Carlo simulation to cover the dynamics of a larger number of vortices. Previous studies using this method mostly sought to explain transport phenomena by examining vortex--defect interactions of moving vortex matter\cite{luo_hu_tvdpn,koshelev_vinokur_vm,fangohr_groot_tvdhf}. Another study examined static, metastable vortex configurations around a defect site\cite{olive_brandt_tvpd}. In our case we are interested in the dynamic behavior and in comparing the simulation to our data.

We employed two different methods for measuring the distortions as described in section II. Section III discusses the results of our 2D simulations showing similar distortions. This gives us confidence that the observations described in section IV are not experimental artifacts.

\section{Distortion detection}

Observing spacial distortions using an STM has to be conducted carefully to ensure that the observations are not just due to piezo artifacts. In our case, we measure lateral distortions on the order of 2~nm using STM images with a pixel resolution of typically $(400\ \mathrm{nm})/(128\ \mathrm{pixel}) = 3.125$~nm/pixel. The computer algorithm to detect the vortex position uses on average 77 pixels per vortex. That reduces the uncertainty by a factor of 8.8 to $0.36$~nm.

We used two methods to measure the distortions within the vortex lattice. In both cases, we first determined the spacial relation for each vortex within one frame. Assuming a triangular lattice, we found the six nearest neighbors for each vortex noting the vortex--vortex distance. The lattice was then pruned by excluding vortices as possible neighbors at distances larger than 1.5 times the average lattice constant. This leaves some vortices, especially close to the edge of the scan area, with coordinations less than six. The local layout is used for the subsequent calculations.

\begin{figure}[tb]
  \begin{centering}
  \includegraphics[width=.495\linewidth]{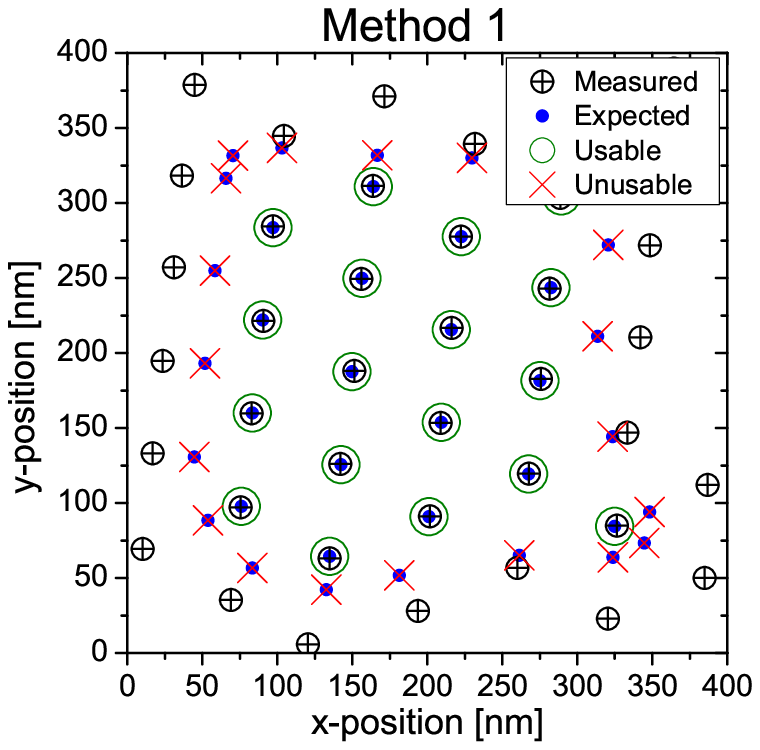}\hfill%
	\includegraphics[width=.495\linewidth]{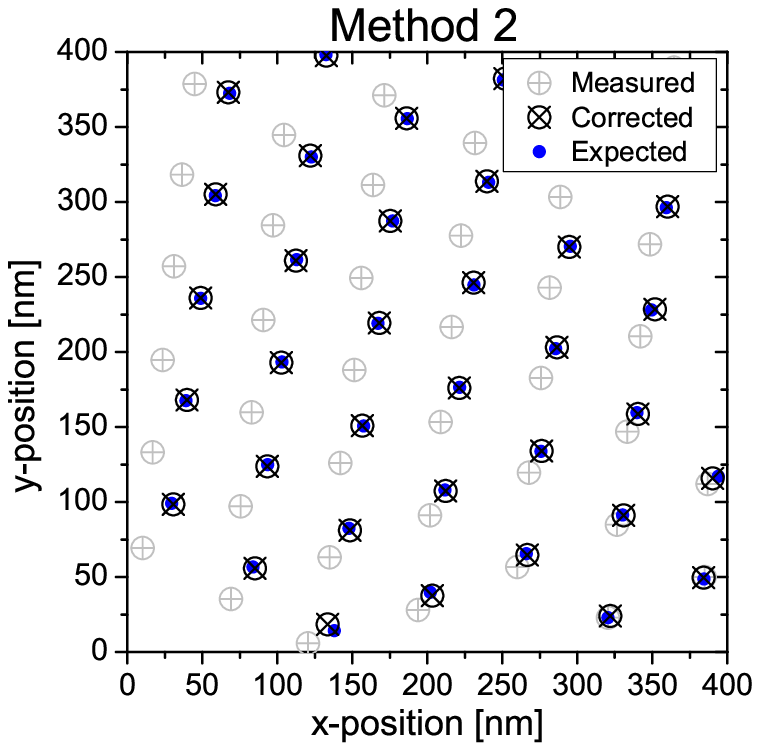}%
	\end{centering}
	\caption{(Color online) Measuring the vortex lattice distortion by means of center of mass (Method 1) and using an ideal lattice (Method 2). Method 1 gives only results for six fold coordinated vortices (green circles) whereas Method 2 allows to measure the distortion up to the edge of the scan area.}
	\label{fig:distdet}
\end{figure}

\begin{figure}[tb]
  \begin{centering}
  \includegraphics[width=.35\linewidth]{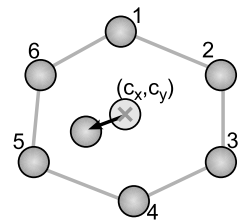}
	\end{centering}
	\caption{Method 1 of the distortion detection. Six neighbours of a vortex form a hexagon (1--6). The center of mass $(c_\mathrm{x},c_\mathrm{y})$ is used as a reference to determine the displacement (arrow).}
	\label{fig:distdet_m1}
\end{figure}

Fig. \ref{fig:distdet} shows both methods applied to the same frame. The first method calculates the lattice distortion as the displacement of a vortex with respect to the center of mass of the hexagon formed by its six nearest neighbors. The center of mass $(c_\mathrm{x},c_\mathrm{y})$ is calculated using a sum formula for an N-sided polygon:
\[ c_\mathrm{x} = \frac{1}{6A}\sum_{i=1}^N (x_i+x_{i+1})(x_iy_{i+1}-x_{i+1}y_i) \]
\[ c_\mathrm{y} = \frac{1}{6A}\sum_{i=0}^N (y_i+y_{i+1})(x_iy_{i+1}-x_{i+1}y_i) \]
with the area \[A = \frac{1}{2}\sum_{i=1}^N (x_iy_{i+1}-x_{i+1}y_i)\]
$(x_i,y_i)$ denotes the $i$th point of the polygon (i.e., measured vortex positions). Fig. \ref{fig:distdet_m1} shows a schematic diagram of the process. This method is easy to implement, but it has two drawbacks: First, if one vortex is displaced, it influences the estimated position for its six neighbors by $\sim\frac{1}{6}$ of its displacement. Second, each vortex has to have six neighbors to be considered (circles in Fig. \ref{fig:distdet}) Consequently, vortices close to the edge of the scan area can not be evaluated (crosses in Fig. \ref{fig:distdet}). Nevertheless, this straight-forward method does not require any further data processing and allows us to validate results obtained by other methods.

The second method is more elaborate. A backtracking algorithm retraces the vortex lattice of a given frame with a fixed lattice constant and orientation yielding a set of expected coordinates $\left\{(e_\mathrm{x},e_\mathrm{y})\right\}$ per image. The measured lattice positions $\left\{(p_\mathrm{x},p_\mathrm{y})\right\}$ are mapped onto the expected coordinates using a 2-D, 2nd order polynomial for the x- and y-coordinate, respectively. The set of parameters of the polynomial $\left\{(a_{j,k},b_{j,k})\right\}$ are estimated by solving:
\[ e_\mathrm{x} = \sum_{j,k\,=0}^2 a_{j,k}\, p_\mathrm{x}^j p_\mathrm{y}^k \qquad\mathrm{and}\qquad
   e_\mathrm{y} = \sum_{j,k\,=0}^2 b_{j,k}\, p_\mathrm{x}^j p_\mathrm{y}^k
 \]
The parameters are then used to map the measured coordinates onto corrected ones $\left\{(c_\mathrm{x},c_\mathrm{y})\right\}$, effectively removing distortions introduced by the scan piezo. The difference between the corrected and expected coordinates shows the lattice distortion: $\vec{d}=\vec{c}-\vec{e}$. This method allows us to consider more vortices and suffers less from back-action of the displacement of some of the vortices.

\begin{table}[tb]
	\centering
		\begin{tabular}{|c|c|c|c|c|c|}
		\hline
		 Series & $N$ & $x$ & $y$ & angle & $|\vec{d}|$\\
		 \hline
     1 & 47091 & 0.86 & 0.87 & 0.76 & 0.75 \\
		 \hline
     2 & 20349 & 0.88 & 0.87 & 0.77 & 0.74 \\
		 \hline
     3 & 44436 & 0.87 & 0.93 & 0.87 & 0.80 \\
     \hline	
		\end{tabular}
	\caption{Correlation of the lattice distortion calculated with two methods within each data set. $N$: number of points used, $x$,$y$,angle,$|\vec{d}|$: correlation of x- and y-component, direction and magnitude of distortion, respectively}
	\label{tab:ComparisonBetweenMethods}
\end{table}
We find a good agreement when comparing the two methods of determining the lattice distortion for three series of measurement. The correlation coefficients are summarized in Table \ref{tab:ComparisonBetweenMethods}. We find a very good correlation for the x- and y-distortion with values well above 0.8. Thus, we focus in the following discussion solely on the data determined by the second method.

\section{Simulation}

\begin{table}[tb]
\begin{tabular}{|l|c|c|}
\hline
Interaction & Force constant [nN/m] & Radius [nm] \\
\hline
Global force & 0.01 -- 20 & -- \\
\hline
 vortex--vortex & 30 -- 40 & 200 \\
\hline
 vortex--defect & -10 -- -1 & 20 -- 120 \\
\hline
\end{tabular}
\caption{Typical force and distance constants used in the simulation.}
\label{tab1}
\end{table}
\begin{figure}[tb]
  \begin{centering}
  \includegraphics[width=.49\linewidth]{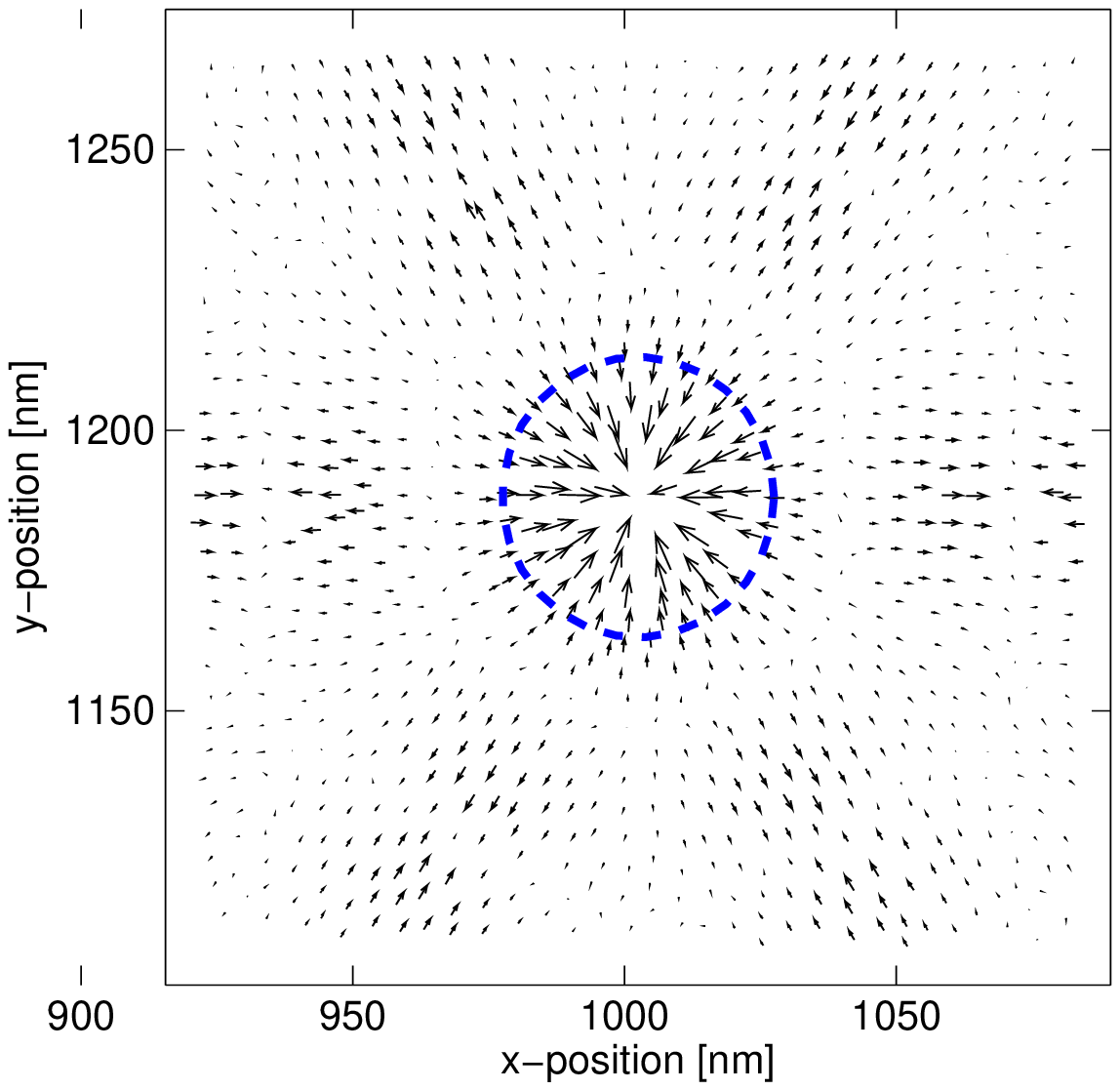}
  \includegraphics[width=.49\linewidth]{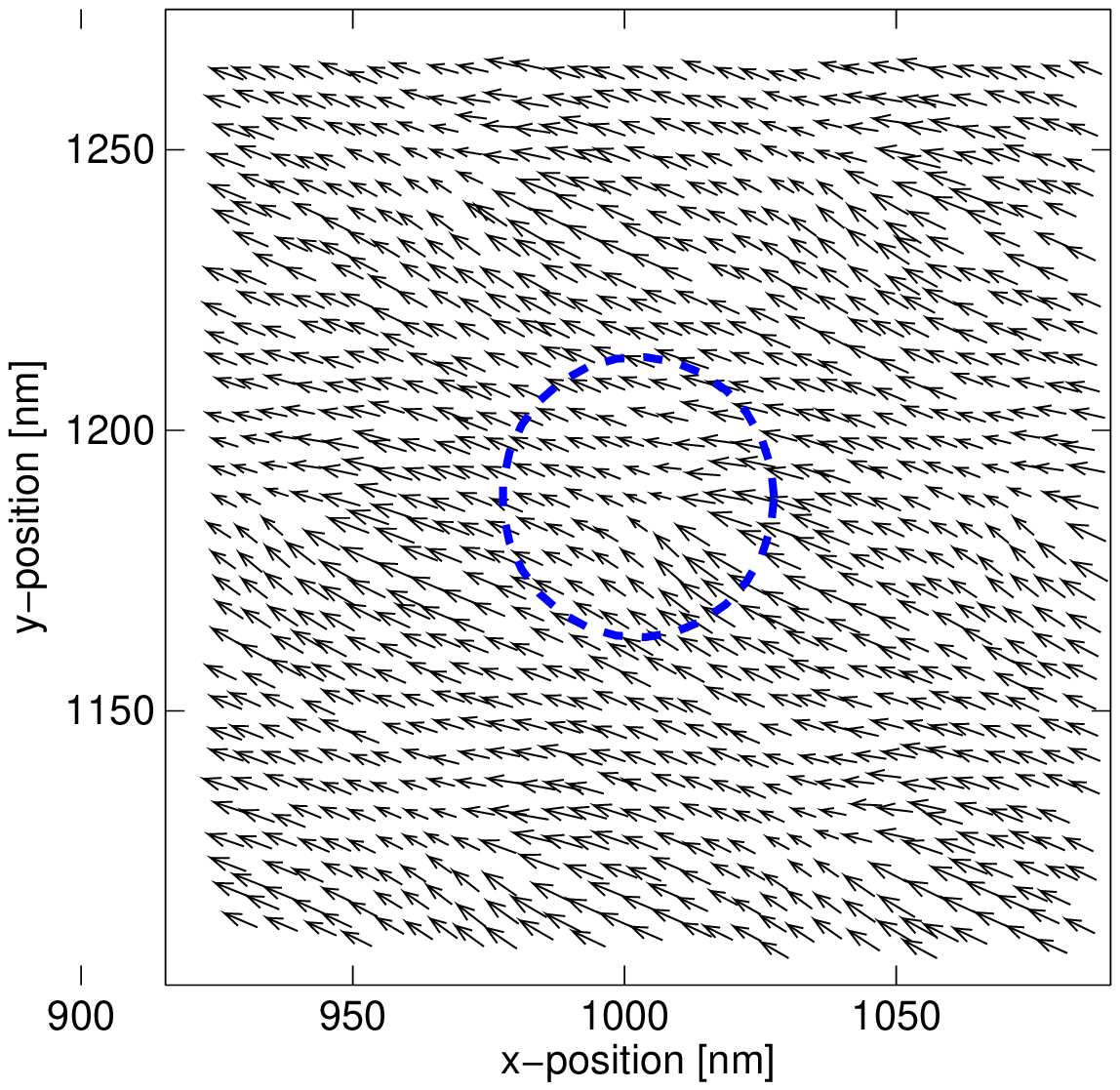}
  \end{centering}
	\caption{(Color online) Lattice distortions (left) and velocity (right) distribution in the vicinity of a single defect. The arrows represent averages of a $5\times5$~nm$^2$ area of about 3 data points each. The scaling factors for the arrows are $m_\mathrm{d}=4$ and $m_\mathrm{v}=0.6 \frac{\mathrm{nm}}{\mathrm{pm/s}}$, respectively.}
	\label{fig:simu}
\end{figure}
The simulation of a moving vortex lattice will be discussed elsewhere in greater detail\cite{md_vsim}. Here it primarily serves as a test whether the effect observed in the data is at all feasible in order to rule out strange behavior of the scan piezo, the microscope electronics, or an influence of local topographic features on determining the vortex position. Similar to previous work of others, we calculate the flow of an ensemble of vortices over a static landscape of defects in two dimensions with periodic boundary conditions under a constant driving force. The full vortex--vortex interaction is taken into account. The equation of motion for the $i$th vortex is given by:
\[ \eta \vec{v}_i = \sum_{j \ne i} \vec{F}_\mathrm{V}\left(\vec{x}_i-\vec{x}_j\right)
                    + \sum_{j} \vec{F}_\mathrm{D}\left(\vec{x}_i-\vec{x}_{\mathrm{D},j}\right)
                    + \vec{F}_\mathrm{global}
\]
Here \[ \vec{F}_\mathrm{V}(\vec{r})=f_{\mathrm{V},0} \cdot \frac{\vec{r}}{|\vec{r}|} \cdot K_1(|\vec{r}|/r_{\mathrm{V},0})\] and \[\vec{F}_\mathrm{D}(\vec{r}) = f_{\mathrm{D},0} \cdot \frac{\vec{r}}{r_{\mathrm{D},0}} \cdot \mathrm{e}^{-(|\vec{r}|/r_{\mathrm{D},0})^2}\] are the vortex--vortex \cite{olive_brandt_tvpd} and vortex--defect interactions, respectively. $K_1$ is the modified Bessel function of the third kind, first order. $\vec{F}_\mathrm{global}$ is the global driving force. The damping of the vortex motion $\eta$ is set to 1 (critical damping). Table \ref{tab1} summarizes typical values for the force and distance constants.

\begin{figure}[tb]
  \begin{centering}
  \includegraphics[width=.8\linewidth]{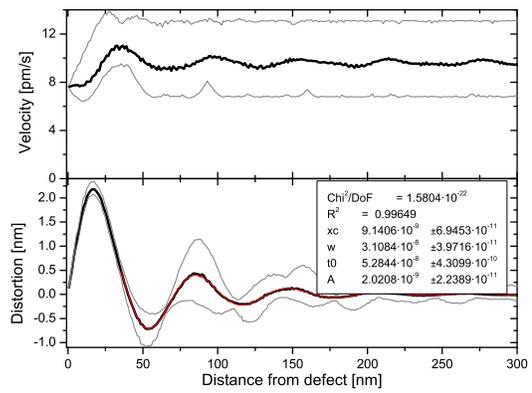}
  \end{centering}
	\caption{(Color online) Distance dependence of the radial component of the lattice distortion and the magnitude of the velocity next to a defect. The plot the average over a 1.2 nm wide ring and time at a given distance as well as the upper/lower bound. An exponentially decaying function was fit to the lattice distortion.}
	\label{fig:simu_dist_dep}
\end{figure}
To test the simulation for lattice distortion we put a single defect within a toroidal area of $2\times2$~$\mu\mathrm{m}^2$. The pinning force had a maximum value of -1~nN and a gaussian profile with a width of 25~nm. The driving force had a value of 10~pN leading to an average velocity of 9.8~pm/s. The vortex density was chosen to be $\sim 242$~$\mu\mathrm{m}^{-2}$ (nearest neighbor distance: $\sim 69$~nm $\Rightarrow B=0.53$~T), similar to the vortex density within the data series considered in this paper.
The simulations show similar velocity and distortion patterns as the measurements. Fig.\ \ref{fig:simu} shows a map of the lattice distortions and the velocity around the defect. The distortion clearly reflects the attractive nature of the defect since all nearby vortices are deflected inward. Additional disturbances with decreasing magnitude, showing roughly the lattice spacing and orientation, can be found farther away. The velocity, however, mostly shows a repeating pattern of variation without a clear indication of the defect position.

\begin{figure}[tb]
  \begin{centering}
  \includegraphics[width=.8\linewidth]{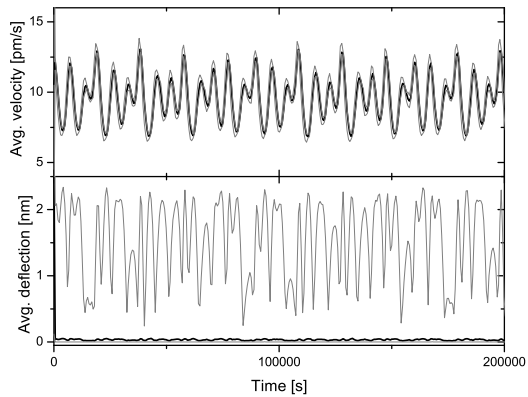}
  \end{centering}
	\caption{Time dependence of the magnitude of the lattice distortion and velocity. The plot shows the upper/lower bound as well as the average. The periodicity in the average velocity is largely due to the periodic boundary conditions.}
	\label{fig:simu_t_dep}
\end{figure}

The different influence of a point defect on the distortion and velocity is also found in distance and time dependent plots (Fig.\ \ref{fig:simu_dist_dep} and \ref{fig:simu_t_dep}). The plots show traces of the average values as well as the envelope. The single defect influences the velocity almost uniformly throughout the simulated area leading to a modulation of the velocity in the time domain. In contrast, significant distortions are only found within $\sim 3$ lattice constance around the defect, while in the time domain the average distortion value is almost zero. Only the envelope indicates distortions of up to 2 nm, but without showing a regular pattern.

To find the decay rate of the distortion with distance, we fit an exponentially decaying periodic function: \begin{equation} f(x) = A \cdot \mathrm{e}^{-\frac{x}{t_0}} \cdot \sin\left( \pi \cdot \frac{x-x_c}{w} \right)\label{equ:fit} \end{equation} to the distortion in Fig.\ \ref{fig:simu_dist_dep} using the least--$\chi^2$--method. The parameters are given in the figure. The fit yields a period of $2\cdot w=62$~nm and a decay rate of $t_0=53$~nm. The slight deviation of the periodicity is probably due to the attractive character of the defect. The decay rate is apparently not in an easy way related to the parameters of the simulation.

\section{Experimental Results and Discussion}

Distortions of the vortex lattice were observed in all $\sim 50$ time series we measured over the course of one year using 5 different NbSe$_2$ samples at a temperature of 4.2~K. The samples were about 5 mm in diameter and up to $t \le 0.5$ mm in thickness. They were cleaved at a pressure $p \le 10^{-7}$ mbar before being introduced into the STM mounted inside the Helium cryostat. After the coarse approach the tip is roughly located in the center of the sample. Although we do not have optical access to the microscope, the distance of the tip from the center of the sample can be estimated. If we assume an expanding vortex lattice as the sole driving force, the distance to the expansion center $r$ is given by $r=\frac{v}{\mathrm{d}a/\mathrm{d}t}\cdot a$. $a$ denotes the average lattice constant, $v$ the average velocity. The value of $r$ is typically below 1 mm.

\begin{figure}[tb]
  \begin{centering}
  \includegraphics[width=.95\linewidth]{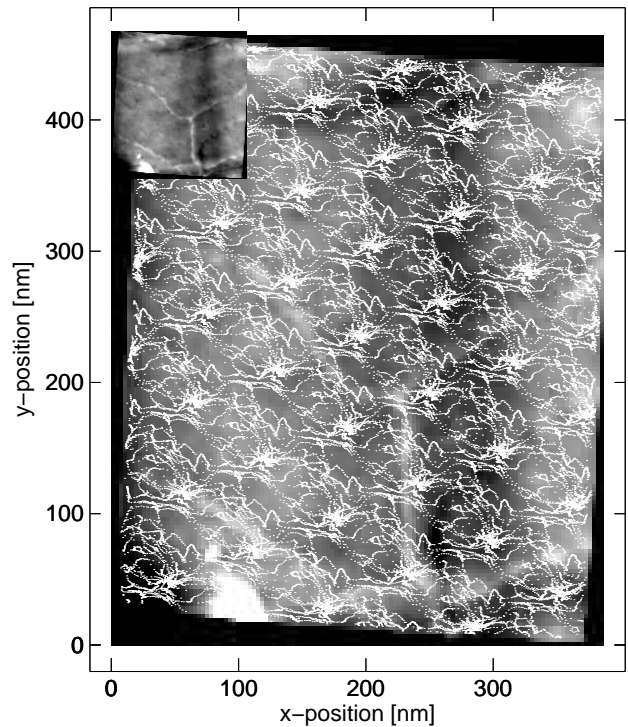}
  \end{centering}
	\caption{Topography overlayed with the track patterns of the vortices passing through the area of three consecutive time series. The inset replicates the topography for clarity. The lines in the topography are single layer steps of NbSe$_2$ (height: 1.25 nm).}
	\label{fig:topotracks}
\end{figure}
 In this paper we focus on three consecutive time series taken within the same area of a sample (here: $r=0.15$ mm). The initial magnetic field of 0.5~T was stored in the superconducting magnet after introducing the sample. The initial relaxation of the vortex lattice leads to an exponential decline of the average velocity with a decay constant of $\sim 30$ min. After a few hours a base velocity of in the order of pm/s remains due to a slow decay of of the magnetic field. The average vortex density shows a slight decline in magnetic field from 0.500~T to 0.497~T during the first time series with a finally value of 0.490~T after 17 days. The average lattice constants and standard deviations are $69.2 \pm 1.6$ nm, $69.4 \pm 1.5$ nm and $69.8 \pm 1.7$ nm, respectively. In between each series, the tip was withdrawn from the sample surface in order to refill the liquid Helium dewar. Topographic images confirm that the location is identical to within one pixel ($=3.125$~nm). Fig.\ \ref {fig:topotracks} shows the topography as well as the combined vortex tracks. Several single steps are visible as well as a particle in the lower left corner and a depression running down the right side. Depressions like this one are regularly observed on layered materials. The tracks run through cluster points reproducing the vortex lattice. The points are connected by seemingly random tracks due to the complexity of the underlying defect pattern. However, each time all vortices follow a particular track owing to the strong vortex--vortex interaction. The later is also responsible for the high level of order in the vortex system. Although the vortex lattice should break up into domains we only observed six fold coordination. Calculating the distance dependent pair correlation function\cite{sow_pair} shows a constant behavior with $B(r)\sim 5.9\cdot 10^{-4}$. This is not entierly unexpected since we are only observing a small area of the vortex lattice of $1.6\times 0.4$ $\mu$m$^2$ --- including the lattice motion.

\begin{figure}[tb]
  \begin{centering}
  \includegraphics[width=.95\linewidth]{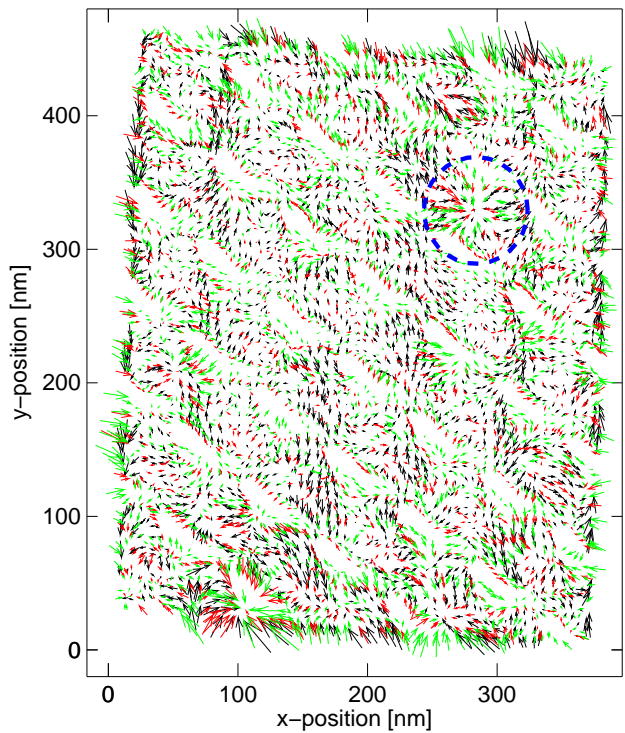}
	\end{centering}
	\caption{(Color online) Overview of the lattice distortion of three consecutive time series. The arrows represent the average distortion within an $8\times8$~nm$^2$ area. The three data series are plotted in different shades of gray. The circle marks an attractive site for the vortices.}
	\label{fig:overview}
\end{figure}
It is still possible to pick out and study single defects by looking at the local distortion patterns. Fig.\ \ref{fig:overview} shows an overview of the lattice distortion within the field of view. Each arrow represents the average distortion of an $8\times8$~nm$^2$ area within the respective data series and points from the `expected' to the `corrected' position. The length of the arrows has been increased by a factor of 4 to improve visibility. Generally, the three data series show a good agreement, but there are a few variations. To test the similarity we calculated the correlation of the local lattice distortion between data sets. The results are presented in table \ref{tab2}. For the lattice distortion the values show a very high correlation ($>0.8$) when comparing series 1 and 2 and still a good agreement ($>0.5$) for series 1 vs. 3 and 2 vs. 3. In comparison the velocity shows little or no correlation. The differences might be explained by the slight increase in the lattice constant leading to a changing interaction with the defect pattern. It is noteworthy that the steps visible in the topographic images do not lead to any discernable effect on the vortex lattice. Previous studies using the Bitter technique\cite{pardo_bishop_evlm,fasano_sup_pin} or scanning SQUID microscopy \cite{Plourde_ss_squid} found clear effects due to steps. However, the studies where carried out at much lower magnetic fields (up to 3.6 mT) leading to a softer vortex lattice and for steps of at least 25 nm in height. In our case the step height represents less than 0.01 \% of the sample thickness and is thus unlikely to have a measurable influence.
\begin{table}[tb]
\begin{tabular}{|c|c|c|c|c|c|c|c|}
\hline
 Series & $N$ & $d_\mathrm{x}$ & $d_\mathrm{y}$ & $\sqrt{d_\mathrm{x}^2+d_\mathrm{y}^2}$ & $\tan^{-1}(\frac{d_\mathrm{y}}{d_\mathrm{x}})$ & $v_\parallel$ & $v_\perp$ \\
\hline
1 vs.\ 2 & 1304 & 0.81 & 0.83 & 0.76 & 0.69 & 0.39 & 0.22 \\
\hline
1 vs.\ 3 & 1565 & 0.53 & 0.52 & 0.54 & 0.39 & 0.05 & 0.35 \\
\hline
2 vs.\ 3 & 1431 & 0.61 & 0.57 & 0.58 & 0.44 & 0.18 & -0.01\\
\hline
\end{tabular}
\caption{Correlation between data series for the $50\times60$ ($\sim 8\times8$~nm$^2$) raster. $N$: Sample size, $d_\mathrm{x}$, $d_\mathrm{y}$: distortion in x and y, respectively, and $v_\parallel$ ($v_\perp$): velocity parallel (perpendicular) with respect to average travel direction.}
\label{tab2}
\end{table}

\begin{figure}[tb]
  \begin{centering}
  \includegraphics[width=.95\linewidth]{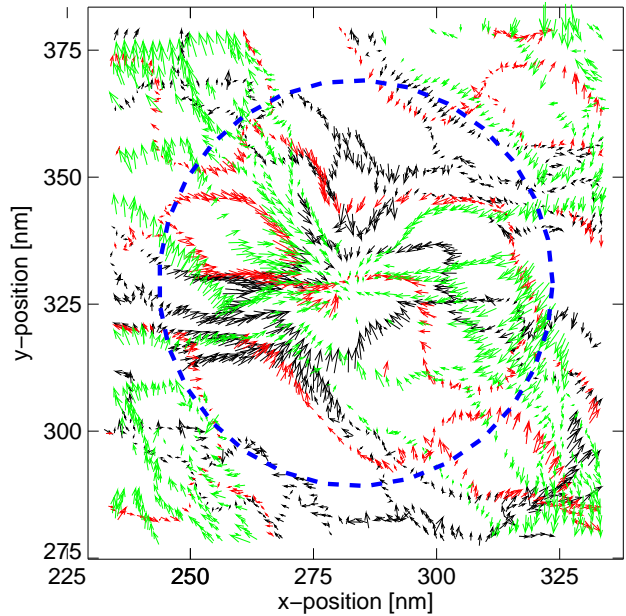}
	\end{centering}
	\caption{(Color online) Detailed view of the distortion pattern around subsurface defect. The arrows represent the average distortion within an $2\times2$~nm$^2$ area. The three data series are plotted in different shades of gray. The arrows within the circle point to a common center. The length of the arrows has been magnified by a factor of 2.}
	\label{fig:detail1}
\end{figure}
A circle in Fig.\ \ref{fig:overview} marks an area were the distortion is relatively high and is apparently pointing to a center position. Similar locations can be found, e.g. in the lower left corner. At the marked position, however, the topography shows no surface features such as steps or particles and was therefore chosen for closer examination. Fig.\ \ref{fig:detail1} shows the same location with a resolution of $2\times2$~nm$^2$. The average number of samples per arrow is 6.5. The length of the arrows has been magnified by a factor of 2. The arrows within the circle point to a common center where we suspect a subsurface defect. We also note, that the defect lies within a cluster point and has a focusing effect as the vortices pass by. The motion occurs essentially from the lower right to the upper left corner. Still, there are variations in the vortex tracks, even in close vicinity of the defect. Therefore, while this pinning center appears to have an influence on the vortex motion, other pinning sites have to be involved to form the track and velocity patterns observed.

\begin{figure}[tb]
  \begin{centering}
  \includegraphics[width=.8\linewidth]{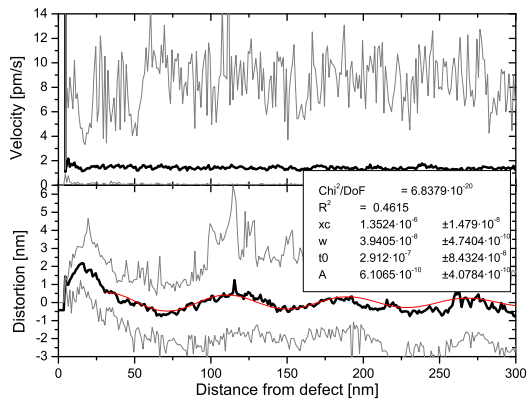}
	\end{centering}
	\caption{(Color online) Dependence of the magnitude of the velocity and the radial component of the lattice distortion on distance from presumed defect location. The plot shows the upper/lower bound as well as the average.}
	\label{fig:mdistdep}
\end{figure} %
\begin{figure}[tb]
  \begin{centering}
  \includegraphics[width=.8\linewidth]{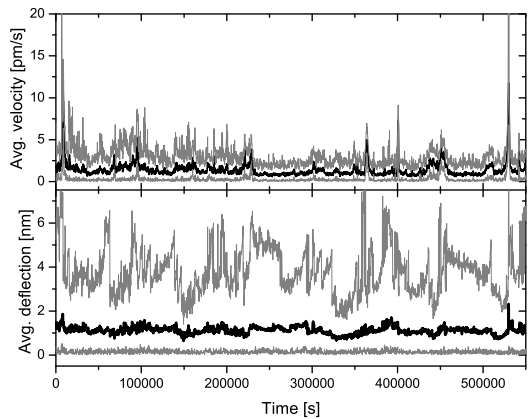}
	\end{centering}
	\caption{Dependence of the magnitude of the velocity and lattice distortion on time. The plot shows the upper/lower bound as well as the average.}
	\label{fig:mtdep}
\end{figure}
Finally, Fig.\ \ref{fig:mdistdep} and \ref{fig:mtdep} show the dependence of the lattice distortion and velocity on the distance from the center found in Fig.\ \ref{fig:detail1} and on time, respectively. The curves show a similar behavior to that found in the simulations (cf. Fig.\ \ref{fig:simu_dist_dep} and \ref{fig:simu_t_dep}). The distance dependence of the distortion shows a clear first maximum and oscillations, while the velocity is featureless. The roles are reversed when analyzing the time dependence. The velocity shows several peaks, stemming from the collective interaction of the vortex lattice with the defects, whereas the magnitude of the average distortion is essentially flat. Fitting equation \ref{equ:fit} to the distance dependence of the lattice distortion yields a period of $2\cdot w=78.8$~nm and a decay rate of $t_0=291$~nm. Given the large width of the envelope after the first maximum and the poor correlation coefficient of $R^2 = 0.46$, one cannot give much weight to these numbers.

\section{Conclusion}

We discovered local lattice distortions in slow moving vortex lattices in NbSe$_2$. The distortions were found in 2D simulations as well as in measurements. It is a local effect generated by the interaction of vortices with individual pinning sites. Thus, observing the distortion patterns allows to identify the location. This is the first step for a detailed study of the vortex--defect interaction. In contrast, variations in the velocity are a collective phenomenon generated through the collective interaction of the vortices with a larger group of defects outside of our observation area, facilitated by the strong vortex--vortex interaction. This becomes evident by examining the distance and time dependence of the distortion and velocity. While the distribution of defects governs the vortex tracks and the patterns in the velocity, the elasticity of the vortex lattice still permits individual vortices to be influenced by individual defects.

\section{Acknowledgment}
We like to thank Prof. Eva Andrei and her group at Rutgers University for supplying NbSe$_2$ samples and for helpful discussions.

\end{document}